\documentstyle[bibnorm]{lamuphys22}
\newcommand{\ft}[2]{{\textstyle\frac{#1}{#2}}}
\begin{document}
\title{Spin dependent D-brane interactions and scattering amplitudes in matrix theory}

\author{J.F.\,Morales\inst{1} \and J.C.\,Plefka\inst{2} \and C.A.\,Scrucca\inst{3}
\and M.\,Serone\inst{4} \and A.K.\,Waldron\inst{5}}

\institute{Department of Physics, University ``Tor Vergata'' of Rome, I-00133 Rome, 
Italy 
\and 
Albert-Einstein-Institut, Max-Planck-Institut f\"ur
Gravitationsphysik, D-14473 Potsdam, Germany
\and
Sektion Physik, Ludwig Maximilian University of Munich, D-80333 Munich, Germany
\and
Department of Mathematics, University of Amsterdam, 1018 TV Amsterdam, The
Netherlands
\and
NIKHEF, P.O. Box 41882, 1009 DB Amsterdam, The Netherlands}

\maketitle

\centerline{Talk given by M. Serone}

\begin{abstract}
Spin interactions between two moving Dp-branes are analyzed using the Green-Schwarz
formalism of boundary states. This approach turns out to be extremely efficient
to compute all the spin effects related by supersymmetry to the leading
$v^4/r^{7-p}$ term. All these terms are shown to be scale invariant, supporting
a matrix model description of supergravity interactions.

By employing the LSZ reduction formula for matrix theory and the mentioned
supersymmetric potential for D0-branes, we compute the t-pole of graviton-graviton
and three form-three form scattering in matrix theory.
The results are found to be in complete agreement with tree level supergravity in 
the corresponding kinematical regime and provide, moreover, an explicit map between 
these degrees of freedom in both theories.
\end{abstract}

\section{Introduction}
The central role played by D-branes \cite{pol} in the description of non-perturbative
phenomena in string theory has motivated in the last three years the study of
their dynamics. In particular, spin dependent long range interactions were first
analyzed in \cite{harv} for the case of D0-branes through duality arguments.
We employ in the following a more direct approach, using the Green-Schwarz 
boundary state formalism \cite{gregut1}, that allows one to study more general brane
configurations \cite{mss1,mss2}. In this approach spin interactions arise by inserting
broken supercharges into the partition function of two branes moving past each
other. Instead of computing the total partition function corresponding to
two moving branes, we can equivalently compute correlation functions of vertex
operators, associated to the velocity, on the world-sheet boundary, together with the
insertion of broken supercharges that encode spin dependencies. \\
Consider now the particular case of two parallel Dp-branes in flat
space-time\footnote{More general configurations can be treated in a similar way.
See \cite{mss2} for details.}.
This system leaves unbroken sixteen supercharges meaning, in the
Green-Schwarz light-cone formalism,
that our action will still admit eight fermionic zero modes $S_0^a$, in the 
notation of \cite{gsw}. Since, as we will see, each insertion of a vertex operator 
associated to the velocity provides at most two of them, it follows that the potential 
between parallel branes starts with $v^4$. On the other hand, the insertion of broken
supercharges can allow non-vanishing results for terms with less powers of $v$, 
providing the lacking fermionic zero modes. Moreover the insertion of 
supercharges induces polarization-dependent 
non-minimal couplings between D-branes, i.e. spin-effects. 
Alternatively, since all these terms are related by 
supersymmetry \cite{harv}, it is natural that they are produced by insertion of 
supercharges.

The correlations we consider, associated to the leading $v^{4-n}/r^{7-p+n}$ terms 
discussed above will be fixed by zero modes only, since all massive non-BPS 
bosonic and fermionic string state contributions will always 
precisely cancel. This implies that these amplitudes, which for large brane 
separations have a clear interpretation as spin-dependent interactions, due to 
supersymmetry, present the same functional form
at {\it all scales} and can be extrapolated to the 
substringy regime where the dominant degrees of freedom
are given by the massless open string states living on the branes. 
This means in particular that a one-loop
matrix theory \cite{bfss} computation should be able to reproduce long range 
spin-dependent supergravity interactions. This has been indeed partially shown in 
\cite{spin}.
Motivated by the argument above, one could also ask
whether matrix theory is able to reproduce supergravity scattering amplitudes.
To date, typical matrix theory computations
involve the comparison of classical gravity source-probe actions with the
background field effective action of super Yang--Mills theory in $(1+0)$ dimensions,
evaluated on straight line configurations \footnote{See 
\cite{banks} for an exhaustive list of references.}.

We will be able to go beyond this aproximation 
by constructing a matrix theory analogue of the Lehmann, Symanzik and Zimmermann (LSZ) 
reduction formula.
In this way we can relate S-matrix elements of asymptotic super-graviton states, whose
explicit form has been worked out in \cite{pw}, to the background field expansion
of the matrix theory path integral. Thereafter, all the dynamics will be encoded
in the matrix theory effective potential. At the one-loop leading level, however,
this potential will precisely coincide with the scale invariant spin dependent D0-brane
interactions computed above through the Green-Schwarz boundary state.
Combining then this effective potential with the knowledge of the asymptotic
super-graviton states in matrix theory, we will be able to extract t-poles of 
S-matrix elements, in the kinematical configuration with vanishing 
Kaluza Klein momentum transfer associated to the
circle compactification \cite{psw1,psw2}. 
As particular example, we report here the case of graviton-graviton 
scattering, showing complete agreement with the corresponding tree level supergravity
amplitude\footnote{The explicit computation for three form-three form scattering, again
in complete agreement with supergravity, has been performed in \cite{psw2}.}. 

Throughout the following we work in the
$N=2$ sector of the matrix model, so that we are actually considering the
Susskind finite $N$ generalisation \cite{suss} of the matrix theory
conjecture.

\section{Spin effects in the GS boundary state}
D-branes, as world-sheet boundaries, can be described by suitable closed string states,
called boundary states, that implement the usual Neumann-Dirichlet boundary conditions,
both in the covariant as well as the Green-Schwarz formalism.
In the latter framework, the boundary state describing a single flat D-brane is defined
by the conditions
\begin{equation}
(\partial X^i+M^i_j\bar{\partial}X^j)|B\rangle = 0 
\end{equation}
and the BPS conditions
\begin{equation}
(Q^a + i M_{a\dot{b}}\tilde Q^{\dot{b}})|B\rangle = 0, \ \ 
(Q^{\dot a}+i M_{\dot a b} \tilde Q^{b}) |B\rangle = 0
\label{bps}
\end{equation}
where 
$Q^a = (2 p^+)^{1/2} \oint d\sigma S^a \;,\;
Q^{\dot a} = (p^+)^{-1/2} \gamma^i_{\dot a a}
\oint d\sigma \partial X^i S^a $
are the usual linearly and non-linearly realized light-cone supercharges.
The same of course for the right-moving ones.
$M_{ij},M_{a\dot{a}},M_{\dot{a}a}$ are definite $SO(8)$ matrices
\cite{gregut1,mss1}, depending on the dimensionality of the brane \footnote{In writing
$M_{a\dot{a}}$ we have implicitly chosen to work in the IIA theory, relevant for the
analysis of D-particles.}. 
The solution for $|B\rangle$ turns out to be
\begin{equation}
\label{mom}
|B\rangle =\exp\sum_{n>0}\left({1\over n}  M_{ij}
\alpha^i_{-n}\widetilde{\alpha}^j_{-n} -
i M_{a\dot{a}}S^a_{-n}\widetilde{S}^{\dot{a}}_{-n}\right)|B_0\rangle
\end{equation}
$|B_0\rangle$ being the zero mode part, 
\begin{equation}
\label{zm}
|B_0\rangle= M_{ij} |i\rangle|\widetilde j\rangle
- i M_{\dot a b}|\dot a\rangle|\widetilde b\rangle
\end{equation}
In this gauge, the $\pm$ light-cone directions satisfy automatically Dirichlet
boundary conditions, meaning that the branes we are dealing with are actually Euclidean 
branes. The boundary state associated to moving branes is obtained by simply
boosting the static one. Although in light-cone gauge this procedure
turns out to be problematic, it is possible to overcome this
difficulty by identifying one of the SO(8) transverse directions with the 
time direction \cite{mss1,mss2}. Thereafter one deduces
the corresponding $SO(1,9)$ expressions and performs a double
analytic continuation to the final covariant result.

The
configuration space boundary state $|B,\vec{x}\rangle$ is given by
\begin{equation}
\label{conf}
|B,\vec{x}\rangle =(2\pi\sqrt{\alpha^{\prime}})^{4-p}\int\frac{d^{9-p}q}{(2\pi)^{9-p}}
\,e^{i \vec{q}\cdot\vec{x}}\,|B\rangle\otimes|\vec{q}\rangle
\end{equation} \noindent
with $\langle q|q^{\prime}\rangle
={\rm vol}_{p+1}\,(2\pi)^{9-p}\delta^{(9-p)}(q-q^{\prime})$
and ${\rm vol}_{p+1}$ is the space-time volume spanned by the p-brane.
In this way the static force between two parallel branes is 
\begin{equation}
{\cal A}=\int_0^\infty \!\!dt \,
\langle B,\vec{x}|e^{-2\pi t\alpha^{\prime}p^+(P^--i\partial/\partial x^+)}| B,\vec{y} \rangle
\label{cyli}
\end{equation}
with $P^-=\frac{(p^i)^2}{p^+}+{\rm osc.}$ the light-cone  
Hamiltonian (the term $i\partial/\partial x^+$ just implements
the $p^-$ subtraction needed to obtain the effective Hamiltonian in  
this gauge). From (\ref{cyli}) we get
\begin{equation} 
{\cal A}=V_{p+1}\,(4\pi^2\alpha^{\prime})^{4-p}\int_0^\infty \! dt 
\int\frac{d^{9-p}q}{(2\pi)^{9-p}}\,e^{i \vec{q}\cdot (\vec{x}-\vec{y})}
\,e^{-\pi t\alpha^{\prime}\vec{q}^2} (8-8)\prod_{n=1}^{\infty}\frac{(1-e^{-2\pi tn})^8}
{(1-e^{-2\pi tn})^8} \label{static}
\end{equation}
where the factor $(8-8)$ is due to the trace performed on the zero mode part of the
boundary state, eq.(\ref{zm}). Note in particular that
massive string contributions precisely cancel from the amplitude.

A generic one-loop $n$-point function of vertex  operators $V_1,\ldots,V_n$
will be then given by
\begin{equation}
{\cal A}_n=\int_0^\infty \!\!dt \,
\langle B,\vec{x}|e^{-2\pi t\alpha^{\prime} p^+(P^--i\partial/\partial
x^+)}V_1\dots V_n
| B,\vec{y} \rangle \label{cylin}
\end{equation}
In particular, by inserting the boost operator
$e^{V_B}$, where \footnote{The direction 1 entering in (\ref{boost}) will
be Wick rotated to give the time direction.}
\begin{equation}
V_B=v_i\oint_{\tau=0}\!d\sigma\left(X^{[1}\partial_{\sigma}X^{i]}
+\frac{1}{2}S\,\gamma^{1i}S\right) \label{boost}
\end{equation}
one recovers the eikonal scattering between moving branes performed
in \cite{bach}.
Dp-branes correspond however to solitonic BPS saturated solutions of 
Type IIA(B) supergravity, which preserve one half of the supersymmetries.
The remaining half is realized on a short-multiplet containing 256 p-brane 
configurations; all the various components of the short-multiplet are related by 
supersymmetry transformations generated by the 16 broken supercharges.
{}From this perspective, the original computation by Polchinski \cite{pol}
or the eikonal scattering \cite{bach} correspond simply to the
leading interaction between two generic components of the
super-multiplet, that does not depend on the particular polarization
of both states \cite{mss1}. In order to get the spin-dependent part
of D-brane interactions, one has to insert broken supercharges
into the amplitude (\ref{cyli}). In particular the correlator
that will encode the eikonal scattering of two moving D-branes,
including all spin effects, is the following:
\begin{equation}
{\cal V}=\frac{1}{2}\int_0^\infty \!\!dt \,
\langle B,\vec{x}=0|e^{-2\pi t\alpha^{\prime}
p^+(P^--i\partial/\partial x^+)}e^{V_B}
e^{(\eta
Q_-+\widetilde{\eta}\widetilde{Q}_-)}|B,\vec{y}=\vec{b}
\rangle 
\label{corr}
\end{equation}
$Q_-,\widetilde{Q}_-$ being the SO($8$) supercharges broken by the  
presence of D-branes (with $\eta, \widetilde{\eta}$ the corresponding supersymmetry
parameters) and $V_B$ the boost operator (\ref{boost}) given above.
Although the full computation of (\ref{corr}) will be extremely
complicated, we will see that the leading interaction terms in
(\ref{corr}) can be easily extracted.
Since a configuration of parallel branes preserves 1/2 of the
supercharges, in light-cone gauge this implies that among the 16 linearly 
realized supercharges $S^a_0,\widetilde{S}_0^{\dot{a}}$, eight of them are 
left unbroken. Equations (\ref{cyli}) and (\ref{corr}) require  then the insertion  
of at least eight zero modes (that, due to the constraints (\ref{bps}), 
can be always chosen to be $S_0^a$)
in order to get a non-vanishing result.  This is better seen in a
path-integral approach: in this case we will have an action admitting eight
zero modes $S_0^a$; unless we do not insert vertex operators
that soak up these zero modes, the Grassmanian integration over them will
always make the amplitude vanishing.

This is indeed a very easy way to understand why the interaction
between parallel moving branes start with the fourth power of velocity;
since $V_B$ can provide at most two zero modes $S_0^a$, the first non-vanishing
correlator has to contain four $V_B$'s. In this particular case, moreover, 
the only role of the vertex operators is to provide these fermionic zero modes.
This implies that all the massive string contributions, being unaffected
by the insertion of the vertex operators, will precisely cancel, exactly as in
the evaluation of the static force (\ref{static}).
These amplitudes are therefore {\it scale invariant}, in the 
sense that their dependence on the D-brane distance $\vec b$ is exact,  
keeping the same functional form at any distance. 
The same argument implies, of course, that all the terms in (\ref{corr})
involving a maximum of eight fermion fields $S^a$ will be scale invariant
and fixed by zero modes.
Since each pair of supercharges in (\ref{corr}) provides two zero modes $S_0^a$
and an extra power of transfer momentum $q$, the leading
effective potential between two moving D-branes will have the following schematic
form, in configuration space:
\begin{equation}
\Gamma \sim \frac{v^4}{r^{7-p}}+\frac{\theta^2 v^3}{r^{8-p}}+
\frac{\theta^4 v^2}{r^{9-p}}+\frac{\theta^6 v}{r^{10-p}}+\frac{\theta^8}{r^{11-p}}
\label{vsch}
\end{equation}
where $\theta=(\eta,\widetilde{\eta})$. This is indeed the form expected for 
spin-dependent interactions between D0-particles \cite{harv}.
In this last case, by expanding (\ref{corr}) and performing all the algebra, one
finds (normalizing to one the $v^4$ term and setting $\alpha^{\prime}=1$)
\begin{eqnarray}
\Gamma_{(1)}(\vec{v},\vec{b},\theta)&=&\Bigl [\, v^4 + 2i\,v^2 v_m
(\theta\gamma^{mn}\theta)\,\partial_n
-2v_p\,v_q (\theta\gamma^{pm}\theta) (\theta\gamma^{qn}\theta)
\,\partial_m \partial_n \nonumber \\
&&\quad - \frac{4i}{9} v_i (\theta\gamma^{im}\theta)
(\theta\gamma^{nl}\theta)
(\theta\gamma^{pl}\theta)\,\partial_m \partial_n \partial_p
\label{POT} \\
&&\quad + \frac{2}{63}(\theta\gamma^{ml}\theta) (\theta\gamma^{nl}\theta)
(\theta\gamma^{pk}\theta)(\theta\gamma^{qk}\theta)
\,\partial_m \partial_n \partial_p \partial_q \Bigl ] \,
\frac{1}{\vec{b}^7} \nonumber
\end{eqnarray}
This is the full one-loop leading potential between two parallel D0-branes,
including their spin interactions; in writing (\ref{POT})
we neglect possible contact terms that are anyway not detectable in this
configuration. Being fixed by the ground states of the Green-Schwarz string only, 
the potential above should be reproduced in particular by a one-loop 
matrix theory computation. Indeed, all the spin interactions computed up to now
in matrix theory \cite{spin} are reproduced by (\ref{POT}).

\section{Super-graviton scattering in matrix theory}
In this section we will show how to compute scattering amplitudes in matrix
theory by using the potential (\ref{POT}) above, the explicit
asymptotic particle states found in \cite{pw} and the Lehmann, Symanzik
and Zimmermann (LSZ) reduction formula applied to matrix theory \cite{psw1,psw2}.
As mentioned in the introduction, we will work in the $N=2$ sector of 
the theory.

The $N=2$ U(2) matrix theory Hamiltonian 
\begin{equation}
H= \ft 12 P^0_\mu P^0_\mu + \Bigl ( \ft 12 \vec{P}_\mu \cdot \vec{P}_\mu
+ \, \ft 14 (\vec{X}_\mu \times \vec{X}_\nu)^2
+ \, \ft i 2 \vec{X}_\mu\cdot \vec{\theta}\, \gamma_\mu\times
\vec{\theta}\Bigr )\, 
\label{MTHam}
\end{equation}
is a sum of an
interacting SU(2) part  describing relative motions
and a free U(1) piece pertaining to the
centre of mass\footnote{In (\ref{MTHam}) a vector notation for the adjoint 
representation of SU(2) is used, $\vec{X}_\mu=(Y^I_\mu,x_\mu)$ and
$\vec{\theta}=(\theta^I,\theta^3)$ ($I=1,2$and $\mu=1,\ldots ,9$).
In the following we will work in a gauge in which $Y^I_9=0$.}. 
The model has a potential with
flat directions along the Cartan directions $x_\mu$ and $\theta^3$, whereas
the remaining degrees of freedom are represented by
supersymmetric harmonic oscillators $Y^I_\mu$ ($\mu\neq9$) and $\theta^I$.
Upon introducing a large gauge invariant
distance $x=(\vec{X}_9\cdot\vec{X}_9)^{1/2}=x_9$ as the separation of 
a pair of particles, the Hamiltonian (\ref{MTHam}) was shown \cite{pw} to
possess asymptotic two particle states of the form 
\begin{equation}
|p^1_\mu,{\cal H}^1;p^2_\mu,{\cal H}^2\rangle=|0_B,0_F\rangle\,
\ft{1}{x_9}e^{i(p^1-p^2) \cdot
x}e^{i(p_1+p_2)\cdot X^0}
|{\cal H}^1\rangle_{\theta^0+\theta^3}\,|{\cal H}^2
\rangle_{\theta^0-\theta^3}\label{state}
\end{equation}
where $|0_B,0_F\rangle$ is the ground state of the superharmonic
oscillators and $p^{1,2}_\mu$, ${\cal H}^{1,2}$ are respectively 
the momenta and polarizations of the two particles.
The polarization states above have been explicitly
constructed in \cite{pw} where it has also been shown how they
fit into the $\underline{44}\oplus\underline{84}
\oplus\underline{128}$ representations of SO(9), corresponding respectively
to the graviton, three-form tensor and gravitino states.

In this hamiltonian approach to matrix theory, once we know the form
of the asymptotic particle states (\ref{state}), it is straightforward to 
form S-matrix elements of the form
$
S_{fi}\, =\, \langle {\rm out}| \exp \{-iHT\} |{\rm
in}\rangle \, 
$.
The object of interest is then
\begin{equation}
{}_{x^\prime_\mu}\langle 0_B,0_F|  \exp \{-iHT\} | 0_B,0_F\rangle_{x_\mu}=
e^{i\Gamma(x'_\mu,x_\mu,\theta^3)}
\label{trans}
\end{equation}
where we have explicitly shown that the ground states $| 0_B,0_F\rangle$ depend on 
the particular vacuum expectation value given to the Cartan moduli 
$x_\mu$ and $x'_\mu$. In general the effective potential $\Gamma$ in (\ref{trans})
can also have a dependence on the fermionic variable $\theta^3$, but not
on the fermionic U(1) term $\theta^0$ that decouples from the interaction.
The vacuum to vacuum transition amplitude (\ref{trans}) may be now
represented as a path integral with appropriate boundary conditions for
the cartan variables
\begin{equation}
e^{i\Gamma(x_\mu,x_\mu^{\prime},\theta^3)}=
\int_{{\vec{X}}_\mu=(0,0,x_\mu),\, {\vec{\theta}}=(0,0,\theta^3)}
^{{\vec{X}}_\mu=(0,0,x_\mu'),\, {\vec{\theta}}=(0,0,\theta^3)}
{\cal D}(\vec{X}_\mu,\vec{A},\vec{b},\vec{c},\vec{\theta})\,
\exp(i\,\int_{-T/2}^{T/2}L_{\rm SYM}).
\end{equation}
The Lagrangian $L_{\rm SYM}$ is that of a supersymmetric
Yang--Mills quantum mechanics with appropriate gauge fixing 
to which end we have introduced ghosts $\vec{b}$, $\vec{c}$ 
and the (Lagrange multiplier) gauge field $\vec{A}$.
The crucial observation is that one can now compute the path integral
above in the gauge of one's choice; in particular it can be computed 
via an expansion about classical trajectories
$X^3_\mu(t)\equiv x_\mu^{\rm cl}(t)
=b_\mu+v_\mu t$ and constant $\theta^3(t)=\theta^3$
which yields the quoted boundary conditions
through the identification $b_\mu=(x'_\mu+x_\mu)/2$ and
$v_\mu=(x'_\mu-x_\mu)/T$.

Up to an overall normalization ${\cal N}$, our LSZ
reduction formula for matrix theory gives
\begin{eqnarray}
S_{fi}&=&\delta^9(k'_\mu-k_\mu)e^{-ik_\mu k_\mu T/2}\nonumber\\
&&\hspace{0cm}
\int d^9x' d^9x \,
\exp(-iw_\mu x'_\mu +iu_\mu x_\mu)
 \langle {\cal H}^3| \langle {\cal H}^4|e^{i\Gamma(v_\mu,b_\mu,\theta^3)}
 |{\cal H}^1\rangle |{\cal H}^2\rangle
\label{superS}
\end{eqnarray}
The leading factor expresses momentum conservation for the centre of mass,
where $k_\mu=p_\mu^1+p_\mu^2$ and $k'_\mu=p_\mu^3+p_\mu^4$ for the in and 
outgoing particles,
respectively, and similarly for the relative momenta
$u_\mu=(p_\mu^1-p_\mu^2)/2$ and $w_\mu=(p_\mu^4-p_\mu^3)/2$.

In a loopwise expansion of the matrix theory path integral one finds
$\Gamma(v_\mu,b_\mu,\theta^3)=v_\mu v_\mu T/2+ \Gamma_{(1)}
+\Gamma_{(2)}+\ldots$ of which we consider only the first two terms
in order to compare our results with tree level supergravity.
Inserting this expansion into~(\ref{superS}) and
changing variables $d^9x' d^9x \rightarrow d^9 (Tv) d^9 b$,
the integral over $Tv_\mu$ may be performed via stationary phase.
Dropping the normalization and the overall centre of mass piece  
the $S$-matrix then reads
\begin{equation}
S_{fi}=e^{-i[(u+w)/2]^2 T/2}
\!\int\! d^9b \,
e^{-i q_\mu b_\mu}\,
 \langle {\cal H}^3| \langle {\cal H}^4|
e^{i\Gamma_{\rm (1)}(u_\mu+w_\mu/2,b_\mu,\theta^3)}
 |{\cal H}^1\rangle |{\cal H}^2\rangle 
\label{sfi}
\end{equation}
where $q_\mu=w_\mu-u_\mu$. It is important to note that in~(\ref{sfi}) the variables
$\theta^3$ are operators
$\{\theta^3_\alpha,\theta^3_\beta\}=\delta_{\alpha\beta}$
whose expectation between polarization states
$|{\cal H}\rangle$ yields the spin dependence of the scattering amplitude.
The loopwise expansion of the
effective action should be valid for the eikonal regime, i.e. large
impact parameter $b_\mu$ or small momentum transfer $q_\mu$.
This is actually the same kinematical regime in which we can trust the potential
(\ref{POT}).

We have now all the tools needed to compute t-poles of
scattering amplitudes in matrix theory. Although one could in this way analyze
arbitrary tree level processes we will consider here graviton-graviton 
scattering; by taking the quantum mechanical expectation value of (\ref{POT}) between
the polarization states in (\ref{sfi}) associated to gravitons, we get the
following t-pole amplitude:
\begin{eqnarray}
{\cal A}&\,\,= \,\,\frac{1}{\textstyle q^2}\,\,\Biggl\{\,\,& 
\ft12(h_1 h_4)(h_2 h_3) v^4 
+ 2\Bigr[(q h_3 h_2 v) (h_1 h_4) 
      - (q h_2 h_3 v) (h_1 h_4)\Bigr] v^2 
\nonumber\\&&\hspace{-.23cm}
+  (vh_2v) (qh_3q)(h_1 h_4) 
+  (vh_3v) (qh_2q)(h_1 h_4) 
- 2(qh_2v) (qh_3v)(h_1 h_4) 
\nonumber\\&&\hspace{-.23cm}
- 2 (qh_1h_4v) (qh_3h_2v)
+ (qh_1h_4v) (qh_2h_3v) 
+ (qh_4h_1v) (qh_3h_2v)  
\nonumber\\&&\hspace{-.23cm}
+ \ft{1}{2}\Bigl [(qh_1h_4h_3h_2q)
-  2(qh_1h_4h_2h_3q) 
+  (qh_4h_1h_2h_3q) 
-  2(qh_2h_3q)(h_1 h_4) \Bigr ] v^2 
\nonumber\\&&\hspace{-.23cm}
-  (qh_2v)  (qh_3q)  (h_1h_4)
+  (qh_2q)  (qh_3v)  (h_1h_4)
-  (qh_1q)  (qh_2h_3h_4v) 
\nonumber\\&&\hspace{-.23cm}
+  (qh_1q)  (qh_3h_2h_4v) 
-  (qh_4q)  (qh_2h_3h_1v)
+  (qh_4q)  (qh_3h_2h_1v)
\nonumber\\&&\hspace{-.23cm}
-  (qh_1v)  (qh_4h_2h_3q) 
+  (qh_1v)  (qh_4h_3h_2q) 
-  (qh_4v)  (qh_1h_2h_3q)
\nonumber\\&&\hspace{-.23cm}
+  (qh_4v)  (qh_1h_3h_2q)
+  (qh_1h_4q)  (qh_2h_3v) 
-  (qh_1h_4q)  (qh_3h_2v)          
\nonumber\\&&\hspace{-.23cm}
+\ft18 \Bigl[ 
   (qh_1q)  (qh_2q)  (h_3h_4) 
+2  (qh_1q)  (qh_4q)  (h_2h_3)
+2  (qh_1q)  (qh_3q)  (h_2h_4) 
\nonumber\\&&\hspace{-.23cm}
+  (qh_3q)  (qh_4q)  (h_1h_2) \Bigr]
+ \ft12\Bigl[
     (qh_1q)  (qh_4h_2h_3q) 
-    (qh_1q)  (qh_2h_4h_3q) 
\nonumber\\&&\hspace{-.23cm}
-    (qh_1q)  (qh_4h_3h_2q) 
-    (qh_4q)  (qh_1h_2h_3q)
+    (qh_4q)  (qh_1h_3h_2q)
\nonumber\\&&\hspace{-.23cm}
-    (qh_4q)  (qh_2h_1h_3q) \Bigr]
+ \ft14\Bigl[
   (qh_1h_3q)  (qh_4h_2q) 
+  (qh_1h_2q)  (qh_4h_3q) 
\nonumber\\&&\hspace{-.23cm}
+  (qh_1h_4q)  (qh_2h_3q) \Bigr]
\, \Biggr\}+\,\, \Bigl[h_1 \longleftrightarrow h_2\, , \, h_3 \longleftrightarrow h_4
\Bigr]
\label{Ulle}
\end{eqnarray}
where again we neglected all terms within the curly brackets proportional
to $q^2\equiv q_\mu q_\mu$, i.e. those that cancel the $1/q^2$ pole.

The result above has to be now compared with the t-pole of tree level
graviton-graviton scattering in eleven dimensional supergravity 
(compactified on a circle). Luckily this computation already appeared
in the literature where it has been shown to be dimension independent
\cite{san}. Since matrix theory is formulated in terms of on shell 
degrees of freedom only, namely transverse physical polarizations 
and euclidean nine-momenta, we have just to fix all the gauge freedom
and take the appropriate kinematics.
Going to light-cone variables, 
we take the case of vanishing $p^-$ momentum exchange
\footnote{We
denote $p_\pm=p^\mp
=(p^{10}\pm p^0)/\sqrt{2}$ and our metric convention is
$\eta_{MN}={\rm diag}
(-,+\ldots,+)$.}, that corresponds to our matrix theory computation,
\begin{eqnarray}
p_M^1=(-\ft12\,(v_\mu-q_\mu/2)^2 ,\, 1\, ,
v_\mu-q_\mu/2 )
&\quad& p_M^2=(-\ft12\, (v_\mu-q_\mu/2)^2 ,\, 1\, , 
-v_\mu+q_\mu/2) \nonumber\\
p_M^4=(-\ft12\, (v_\mu+q_\mu/2)^2 ,\, 1\, ,
v_\mu+q_\mu/2) &\quad & p_M^3=(-\ft12\,
(v_\mu+q_\mu/2)^2 ,\, 1\, ,
-v_\mu-q_\mu/2) \nonumber
\end{eqnarray}
where momenta are measured in units of the compactified radius, so that
$p^-=1$. Note that the vectors $u_\mu$ and $w_\mu$ of~(\ref{superS}) are simply
$u_\mu=v_\mu-q_\mu/2$ and $w_\mu=v_\mu+q_\mu/2$.
Polarizations and momenta are subject to the de Donder gauge
$p^i_N h^i{}_M{}^N-(1/2)p^i_M h^i{}_N{}^N=0$ (no sum on $i$).
We reduce to physical
polarizations by using the residual gauge freedom to set $h^i_{+M}=0$
and solve the de Donder gauge condition in terms of the transverse traceless
polarizations $h^i_{\mu\nu}$ for which $h^i_{-M}=-p^i_\nu h^i_{\nu M}$.

Finally, by plugging the above states in the amplitude reported in \cite{san} and
taking the t-pole\footnote{$t=q_\mu^2=-2p^1_M p_4^M$
in the above parametrization.} part of it, one finds precisely the matrix theory 
amplitude (\ref{Ulle}).

Although the agreement above might have been expected from the scale invariance
of the potential (\ref{POT}), it is clear that these results establish a very precise
correspondence between the two models. In particular they pose the basis to analyze
and interpret higher order matrix theory amplitudes that would correspond to loop
effects in supergravity or M-theory. It is clear that such tests will be fundamental
to establish the range of validity of matrix theory.


\end{document}